\documentclass[12pt,dvips]{article}
\usepackage{graphicx}
\newcommand{\bm}[1]{\mbox{\boldmath$#1\!$}}
\newcommand{\bn}[1]{\mbox{\boldmath$#1$}}

\newcommand{\beq}{\begin{equation}}
\newcommand{\eeq}{\end {equation}}
\newcommand{\bea}{\begin{eqnarray}}
\newcommand{\eea}{\end{eqnarray}}

\begin{document}
\lineskip=24pt
\baselineskip=24pt
\raggedbottom
{\Large{\centerline{\bf Atomic reflection off conductor walls}}}
{\Large{\centerline{\bf as a tool in cold atom traps}}}
\vskip 1.0cm
\centerline{{\bf  M. Al-Amri$_{*}^{\dagger}$} and {\bf M. Babiker}$^{\dagger}$}
\centerline{$\dagger$ Department of Physics, University of York, Heslington, York YO10 5DD, England}
\centerline{$*$ Department of Physics, KKU, Abha P O Box 9003, Saudi Arabia}
\vskip 1.0cm
\section*{Abstract}

We explain why a system of cold $^{85}Rb$ atoms at temperatures of the order $T\approx 7.78\times 10^{-5}$ K 
and below, but not too low to lie in the quantum reflection regime, should be automatically repelled from 
the surface of a conductor without the need of an evanescent field, as in a typical atom mirror, to counteract the van 
der Waals attraction.  The
repulsive potential arises naturally outside the conductor and is effective at distances from the
conductor surface of about $400$nm, intermediate between 
the van der Waals and the Casimir-Polder regions of variation.  We propose that such a field-free reflection capability should be useful
as a component in cold atom traps.
It should be practically free of undesirable field
fluctuations and would be operative at distances for which surface roughness, dissipative effects and other finite conductivity 
effects should be negligibly small.

\vskip 1.0cm

\centerline{\bf {PACS Numbers: 34.50.Dy, 32.80.Pj, 42.50.Vk}}

\newpage

There is currently much interest in the physics of cold atomic
gases fuelled, possibly, by the repeated successes of experiments creating Bose-Einstein condensates (BECs) in many laboratories. Advances 
in laser cooling and 
trapping, have enabled dilute atomic gases to be progressively cooled and confined and, ultimately, made to condense under 
appropriate conditions [1]. The technique of evaporative cooling, in particular, has been successful in cooling both
bosonic and fermionic gases, but the attainment of high density BECs possessing long coherence times 
remains one of the primary goals in this field. 

The controlled manipulation of dilute atomic gases invariably requires a practical means to trap them in a suitable region
of space.  It is generally perceived that a material container would be 
inappropriate as a component of an atomic trap.  This rationale behind this suggestion is that atoms approaching the container walls 
would be subject to the infinitely  attractive van der Waals potential well in which they could fall and, so, could become adsorbed to 
the wall.  This mechanism is regarded to be one of the primary sources of loss
limiting the coherence properties of the ensemble. It is generally understood that a repulsive potential is required to counteract
the effects of the van der Waals attraction, as in an atomic mirror. As we suggest here, this is not true in general; it should be possible for atoms in a 
certain energy range to be reflected without the need of an additional repulsive potential.

The experimental techniques for studying the centre of mass motion of atoms near surfaces has advanced considerably in recent years.
An interesting article by Bushev et al. [2] reports the results of a beautiful experiment in which the potential acting on an atom well 
localised in a standing light wave at a mirror has been measured as a function of distance from the mirror. Equally impressive 
experiemental work
has been done by Shimizu [3], Shimizu and Fujita [4], DrUzhinia and DeKieviet [5], and Pasquini et al [6] on the quantum reflection of 
of atoms possessing extremely small velocity components normal to material surfaces.

The primary aim of this article is  the study of reflection of atoms whose normal velocity components are such that the motion is regarded
as classical and so far removed from the regime in which quantum reflection occurs [4-6].  
The physical principles in the classical regime are closely linked to 
those underlying an atom mirror [6-21] - a device which acts to repel atoms from its surface before they become subject to the 
influence of the attractive van der Waals force. A specific type of atom mirror that has been the subject of considerable investigation
is the evanescent mode atom mirror. The repulsive force field which prevents the atoms from falling into the van der Waals potential well
is created here by application of laser light, arranged in such a manner as to give rise to an evanescent field in the vacuum region
outside the surface of a planar dielectric. When the frequency of the light is appropriately detuned from the atomic transition 
frequency, the dipole force field set up by this evanescent light may overcome the attractive van der Waals force between the atom and 
the mirror surface, creating an overall repulsive barrier for the approaching atom.
The higher the barrier,  the more effective is the system as an atom mirror for the reflection of relatively energetic atoms.

The factors which limit the performance of such mirrors stem mainly from heating effects and also from fluctuations arising from the application of light.  These fluctuations would become 
insignificant at low light intensities. A reduction  of fluctuation effects at low intensities was the main reason for investigations to
devise an atom mirror operating at low intensities.  This requires the modification of the main mirror ingredients to achieve an enhancement at low intensities.
A particular method for achieving enhancement is by the addition of a metallic capping layer [22].  It is this arrangement in which as 
metallic surface is a prominent feature that forms the basis of the work in this article.

We begin by focusing on the variations of the full potential acting in conjunction with 
very low evanescent field intensities, and also in the absence of such an evanescent field, on atoms approaching a high conductivity metallic 
surface.  We proceed to demonstrate the implications of this for the repulsion of atoms in a given energy range approaching such a surface. This
will allow us to draw useful conclusions about the containment of cold atoms using metallic boundaries.

A central requirement of the dynamics of atoms  near a material surface is that the correct potential sampled by an approaching atom be defined over the whole range of atomic position
$z$ from the planar surface of a metal.  Two limits of this potential are well known as the van der Waals potential form, applicable 
at short distances $(z \ll \lambda/4\pi)$, and the Casimir-Polder form at large distances $(z \gg \lambda/4\pi)$. Each of these potentials have
been studied extensively by both theory and experiment [23-26].  However, 
as we have just emphasised, the full range, including the intermediate distance range is important in order to account correctly for
the motion of atoms outside the metallic surface.
Unfortunately, however, analytical forms of 
the potential are known only for the van der Waals and the Casimir forms. For an atom in its ground state the 
variations with distance $z$ over the entire 
range $0< z < \infty$ can only be accessed numerically from complicated analytical expressions, Eq. (3) below, which have been 
derived using a fully quantum electrodynamical procedure [27].  The numerical evaluation is necessitated by the need to include contributions to the potential arising, 
in general, from an infinite sum over a complete set of stationary states other than the ground state [27].  
The potential is a linear combination of two contributions, $U^{\parallel}_{g}$ and $U^{\bot}_{g}$ , depending on whether the transition dipole moments are 
perpendicular or parallel to the surface.  We have 
\beq
U^{\bot,\parallel}_{g}(z)=\frac{1}{(2\pi)^{2}\epsilon_{0}}\sum_{j\neq g}
\frac{\left|\left<{\bm {\mu}}_{\bot,\parallel}\right>_{gj}\right|^{2}}{\Lambda_{jg}^{3}}
{\cal F}^{\bot,\parallel}_{jg}\left(\frac{2z}{\Lambda_{jg}}\right)\label{poten}
\eeq
where $g$ stands for the ground state and the sum over $j$ spans all atomic states $j$ (unperturbed energy $W_{j}$) connected to 
$g$ by electric dipole transitions for which the matrix elements are $\left<{\bm {\mu}}_{\bot,\parallel}\right>_{gj}$, where ${\bm {\mu}}_{\bot,\parallel}$ denotes
the dipole moment operator perpendicular, parallel to the interface at $z=0$.  The notation is such that
\beq
\Lambda_{jg}=\frac{\hbar c}{W_{j}-W_{g}}
\eeq
and ${\cal F}^{\bot,\parallel}_{jg}$ are well defined functions of the dimensionless variable $x=(2z/\Lambda_{gj})$, but
are too complicated to be displayed here. 
The numerically evaluated variations of the potential with distance $z$ of $U^{\bot,\parallel}_{g}(z)$ and their sum $U_{g}$ for $^{85}Rb$ atoms are shown in Fig. 1.

In order to determine the effects of the potential on the operation of an evanescent mode atomic mirror, with a metallic capping layer,
including the field-free and low-intensity cases, we need to incorporate the new potential in the mirror theory.
The basic components of the atomic mirror are shown in Fig.2.  Here a metal forms the capping layer on the planar surface of 
a glass substrate of dielectric constant $\varepsilon_{1}$.  Laser light of angular frequency $\omega$ and intensity $I$ strikes the
metal/glass interface at an angle of incidence 
$\phi$, as shown in Fig.2.  This light
is internally reflected, and partially transmitted into the metal.  The light within the metal subsequently generates the 
evanescent light in the vacuum region.  A neutral atom of mass $M$ possessing a transition frequency $\omega_{0}<\omega$ and
moving in the plane of incidence towards the structure would be subject to a repulsive dipole force plus  a dissipative 
force, both associated with the evanescent field.  The atom would, further, be subject to an additional 
force due to its proximity to the metal. 
 
The dynamics of the atom is governed by the equation of motion
\beq
M\frac{d^{2}{\bf r}}{ dt^{2}}=-Mg{\bf {\hat z}}+{\bf  F}_{\rm {rad}}+{\bf F}^{f}\label{newton}
\eeq
subject to given initial conditions.  Here the first term on the right hand side is the gravitational force.  The second term
${\bf  F}_{\rm {rad}}={\bf F}_{s}+{\bf F}_{d}$ is the radiation force, comprising, respectively, the spontaneous force and the dipole
force arising from evanescent light. The final term in Eq.(\ref{newton})
is the force corresponding to the new potential.

Consider first the radiation force ${\bf  F}_{\rm {rad}}$.  In the present context this force can be written as [18]
\beq
{\bf F}_{\rm {rad}}({\bf r,v})=2\hbar\left\{\frac{\Gamma\Omega^{2}{\bf k}_{\parallel}-{\hat {\bf z}}\Omega\Delta d\Omega/dz}
{\Delta^{2}+2\Omega^{2}+\Gamma^{2}}\right\}={\bf F}_{s}+{\bf F}_{d}
\eeq
where ${\bf k}_{\parallel}$ is the in-plane wavevector of the light.  Its magnitude ${k}_{\parallel}$ is 
 given by $c^{2}k_{\parallel}^{2}=\omega^{2}\epsilon\sin^{2}\phi$, where $\phi$ is the angle of incidence,
$\Delta=\omega-\omega_{0}-{\bf k}_{\parallel}.{\bf v}$ where ${\bf v}$ is the velocity vector of the atom; $\Gamma$ is
the spontaneous emission rate of the upper atomic level and finally $\Omega$ is the Rabi frequency of the atom in the evanescent light.
The square of the Rabi frequency can be written as follows
\beq
\Omega^{2}(z)=J(\phi,n, I)e^{-2k_{z}z}
\eeq
where $k_{z}^{2}=k_{\parallel}^{2}-\omega^{2}/c^{2}$. 
Besides the characteristic exponential dependence on the atomic position $z$, the squared Rabi frequency is proportional to the 
factor $J$, which depends on a number of parameters, namely the intensity of the 
light and its angle of incidence at the metal/dielectric interface, as well as the metallic electron density.

Consider next the last term ${\bf F}^{f}$ in Eq.(\ref{newton}) which arises
because of the proximity of the atom to the surface.  This is formally related to a new field-dependent potential $U_{g}^{f}$ by  
\beq
{\bf F}^{f}=-\frac{\partial U_{g}^{f}}{\partial z}
\eeq
The new field-dependent potential is evaluated using the field dipole orientation
picture [28].  In this picture  the transition dipole moment vector for all transitions involving the ground state adjusts its direction
 along the local electric field due to the evanescent light as the atom moves towards the 
metallic surface. At any given point the potential $U_{g}^{f}(z)$ emerges as a sum of contributions involving $U^{\bot}_{g}(z)$  
and $U^{\parallel}_{g}(z)$, as determined by the local field direction.  We have
\beq
U_{g}^{f}=U_{g}^{\bot}(z)\sin^{2}\gamma(z)+U_{g}^{\parallel}(z)\cos^{2}\gamma(z)
\label{dipole-orient}
\eeq
where $\gamma$ is the field orientation angle at the location of the atom, given by
\beq
\gamma(z)=\tan^{-1}\left|\frac{E_{z}}{E_{\parallel}}\right|
\eeq
with $E_{z}$  and $E_{\parallel}$  the field components perpendicular 
and parallel to the surface.
 The distribution of $\gamma(z)$ depends on the  internal angle  $\phi$
of incidence within the dielectric region of the structure [28]. 

In the field-free case the potential is the same as the total potential $U_{g}$ in Fig. 1 
\beq
U_{g}^{0}\equiv U_{g}=U^{\bot}_g(z)+U^{\parallel}_g(z)
\eeq 
where the superscript in $U^{0}_{g}$
signifies the absence of the evanescent field.
It is important to note that the evanescent light influences the motion of the atom in two ways.  Firstly it creates a
dissipative force ${\bf F}_{s}$ which is responsible for the in-plane motion of the atom, and, secondly, it sets up a dipole force 
${\bf F}_{d}=-{\bn {\nabla}}U_{d}$ where $U_{d}(z)$ is the dipole potential, which is repulsive for blue detuning.  
The dipole potential combines with $U^{f}_{g}$ to give a total potential $U_{\rm{total}}$ 
\beq
U_{\rm {total}}=U_{g}^{f}+ U_{d}
\eeq
  
At high intensities,  the structure can act as a mirror for relatively energetic atoms, as depicted in the inset to Fig.3,
showing the trajectory of $^{85}Rb$ atoms for a typical set of parameters. The high intensity mode of operation, however, would be beset by
field  fluctuations, heating and, since the reflection action occurs nearer the surface, the effects of surface imperfections cannot be 
ignored.
 
At low, albeit finite, field intensities, the field-dipole orientation mechanism still forces the induced dipole moment vectors for all 
possible transitions to be aligned adiabatically along the local field direction, but the contribution of the dipole potantial $U_{d}$
to the overall potential is negligibly small, so the mirror action is primarily due to $U_{g}^{f}$.
In the field-free case, the repulsion is solely due to 
$U_{g}^{0}$, which coincides with the situation in Fig.1 in which the total potential is the sum of contributions from dipole components
parallel and perpendicular to the surface.  The rationale for using the sum is that in the absence of field orientation, all transition 
dipole moment vector components
are effective in the coupling to the electromagnetic fields.

Figure 4 displays the variation with distance $z$ of the low intensity field-dependent potential $U_{total}$ and compares this with that 
of the field-free potential $U^{0}_{g}$.  The atomic trajectories, evaluated using Eq.(\ref{newton}) arising from the two potentials 
together with gravity and for the same initial conditions 
are seen to be quite different. The initial value ($v_{z}=-0.067$ms$^{-1}$) of the normal component of the velocity is sufficiently small 
for the atom to be 
reflected at $z\approx 865$nm by the second peak of the field-free potential.  When weak light is applied, however, 
the field dipole-orientation mechanism 
leads to a point by point lower potential than the field-free case.  This is because the evanescent field has a larger field component
parallel to the surface which, by virtue of Eq.(\ref{dipole-orient}), leads to a lower potential.  To be reflected under these conditions,
the atom has to penetrate to the first peak 
closer to the mirror surface ($z\approx 340$nm).
The important point here is that mirror action occurs in the field-free case for atoms at such low initial normal velocities and in 
fact more effectively than in the case of a low intensity evanescent light. Note that the distances at which the reflection occurs (hundereds
of nanometres) implies that the assumption of perfect conductivity for the metal is a good one in this context.  Finite conductivity 
would lead to effects that influence the atomic motion at distances of the order $z\approx c/\omega_{p}$ where $\omega_{p}$ is the 
plasma frequency of the metal. For high conductivity metals this is of the order of a few $\AA$s.

It is therefore reasonable to suggest, in view of the results displayed above, that it should be possible for a container with an inner 
metallic surface to repel cold atoms, provided that their velocity component normal to the surface at any point on 
the container inner boundary does not exceed a certain maximum value, $|v_{z}|^{max}$, depending on the atom. All atoms with normal 
velocities higher than this maximum would fall into the van der Waals image potential well and may be lost by adsorption to the surface.  

For orientation as to orders of magnitude it is seen from Fig. 4 that the positions of the main peaks in $U_{g}^{f}$ and $U_{g}^{0}$ occur at a points $z$ in excess of $300$ nm, 
sufficiently far removed from the metallic surface for surface roughness effects and also the finite conductivity dissipative effects to be negligibly small.  
The energy heights of the main potential peaks are
\beq
U_{g}^{f,max}\approx 0.57\hbar\Gamma_{0};\;\;\;\;U_{g}^{0,max}\approx 1.62\hbar\Gamma_{0}; 
\eeq
where $\Gamma_{0}=6.1\times 10^{6}$s$^{-1}$ for the rubidium transition $\omega_{0}=2.42\times 10^{15}$ s$^{-1}$.  The potential peak 
energy for low intensities
corresponds to a temperature of the order $T\approx 2.6\times 10^{-5}$ K, while for the field-free case it is 
$\approx 7.78\times 10^{-5}$ K. 
A gas of $^{85}Rb$ atoms for which the normal velocity component $v_{z}$ is such that $|v_{z}|
=\sqrt{2kT/M}$ would be reflected from the surface.  For $^{85}Rb$ this is of the order $|v_{z}|^{max}\approx 7.0\;{\rm cm\;s}^{-1}$
for the low-intensity case and $|v_{z}|^{max}\approx 12.1\;{\rm cm\;s}^{-1}$
in the field-free case.

We propose the following scenario applicable specifically to $^{87}$Rb.  An ensemble of such atoms that have been pre-cooled to about 
$\approx 7.78\times 10^{-5}$ K are forced to enter a container (cavity) with perfectly conducting walls with typical dimensions 
in the tens of micrometer
range. Any atom in this ensemble will have velocity components that are at most of the order $12.1\;{\rm cm\;s}^{-1}$
and hence can be reflected at each encounter with the container walls.  All such atoms in the ensemble will therefore be  
forced to spend most of their time in the inner regions of the cavity, reflected inwards, whenever they approach the container boundaries. Further laser cooling of the atoms within the container could then be arranged.
We suggest that this scenario could be exploited for the
containment of cold atoms and the subsequent creation of BECs of high densities and long coherence times.

\subsection*{Acknowledgments:} The authors are grateful to Dr Gediminas Juzeliunas for useful discussions.

\newpage

\section*{\bf References}
\begin{enumerate}

\item L. Pitaevskii and S. Stringari, `Bose-Einstein Condensation' (Clarendon Press, Oxford, 2003)

\item P. Bushev, A. Wilson, J. Eschner, C. Raab, F. Smidt-Kaler, C. Becher and R. Blatt, Phys. Rev. Lett. {\bf 42}, 223602 (2004)

\item F. Shimizu, Phys. Rev. Lett. {\bf 86}, 987 (2001)

\item F. Shimizu and J. Fujita, Phys.  Rev. Lett.  {\bf 88}, 123201 (2002)

\item V. Druzhinina and M. DeKieviet, Phys.  Rev. Lett. {\bf 93}, 223201 (2003)

\item T. A. Pasquini, Y. Shin, C. Sanner, M. Saba, A. Schirotzek, D. E. Pritchard and EW. Ketterle, 
Phys.  Rev. Lett. {\bf 93}, 223201 (2004)

\item  R J Wilson, B Holst and W Allison,  Rev. Sci. Instrum {\bf 70}, 2960 (1999)

\item  D C Lau,  A I Sidrov, G I Opat et al,  Eur.  Phys. J D {\bf 5}, 193 (1999)

\item   R Cote, B Segev and M G Raisen,  Phys.  Rev. {\bf A58}, 3999 (1998)

\item   N Friedman, R Ozeri and N Davidson,  J. Opt. Soc. Am. {\bf 15}, 1749 (1998)

\item   P Szriftgiser, D Guery-Odelin, P Desbiolles et al, Acta Phys. Pol. {\bf 93}, 197 (1998)

\item  L Santos and L Rose,  J. Phys. B - At. Mol. Opt. {\bf 30}, 5169 (1997)

\item   A Landragin, J Y Courtois,  G Labeyrie et al,  Phys.  Rev.  Lett. {\bf 77}, 1464 (1996)

\item   N Vansteenkiste,  A Landragin,  G. Labeyrie et al,  Ann. Phys.-Paris {\bf 20}, 595 (1995)

\item   A Landragin, G. Labeyrie,  J Y Courtois et al,  Ann.  Phys.-Paris {\bf 20}, 641 (1995)

\item    G Labeyrie, A Landragin J. von Thanthier et al, Quantum Semicl.  Opt. {\bf 8}, 603 (1996)

\item  N Vansteenkiste,  R Kaiser,  Y Levy et al,  Ann.  Phys.-Paris {\bf 20}, 129 (1995)

\item  S M Tan and D F Walls,  Phys.  Rev. {\bf A50}, 1561 (1994)

\item  J P Dowling and J Gea- Banacloche,  Adv. Atom. Mol. Opt. Phy. {\bf 37}, 1 (1997)

\item  S Feron,  J Reinhardt,  S Lebouteux et al,  Opt.  Commun. {\bf 102}, 83 (1993)

\item  T Esslinger,  M Weidenm\"{u}ller, A Hammerich and T W H\"{a}nch,  Opt.  Lett. {\bf 18}, 450 (1993)

\item   C. R. Bennett, J. B. Kirk and M. Babiker, Phys. Rev. {\bf A63}, 033405 (2001)

\item H. B. C. Casimir and D. Polder, Phys.  Rev. {\bf 73}, 360 (1947)

\item Y. Tikochinsky and L. Spruch, Phys.  Rev. {\bf A48}, 4213 and 4223 (1993)

\item Z. -C. Yan, A. Delgarno and J. F. Babb, Phys.  rev. {\bf A55}, 2882 (1997)

\item Z. -C. Yan and J. F. Babb, Phys. Rev. {\bf 58}, 1247 (1998)

\item   M. Al-Amri and M. Babiker, Phys. Rev. A69, 065801 (2004)

\item   M. Babiker and S. Al-Awfi, J. Mod. Optic. {\bf 48}, 847 (2001)

\end{enumerate}

\newpage

\section*{Figure Captions}

\subsection*{\bf Figure 1}

Variations with distance $z$ of the model potentials $U_{g}^{\parallel}$  and $U_{g}^{\bot}$ and their sum $U_{g}$ for a $^{85}Rb$ atom  
in vacuum in front of a perfect conductor, evaluated using Eq.(\ref{poten}).
The dotted curve presents the variations of $U_{g}^{\bot}$, the dashed curve $U_{g}^{\parallel}$ and the
dash-dotted curve their sum $U_{g}$.
At very short distances $U_{g}$ follows the van der Waals image potential (shown by the solid curve), but substantial deviations 
occur in the intermediate distance range.

\subsection*{\bf Figure 2}

 Schematic arrangement of an evanescent mode atom mirror with a metallic capping layer.  See the text for definition of the
symbols in this figure.

\subsection*{\bf Figure 3}

 The high intensity case $(I=5.0\times 10^{5}$ Wm$^{-2}$) for a typical mirror arrangement with a metallic capping layer for an atomic
transition of energy $\hbar\omega_{0}=1.5$ eV and blue detuning $\Delta=500\Gamma_{0}$, where $\Gamma_{0}=6.1\times 10^{6}$s$^{-1}$. 
 The main figure shows variations (dotted curve) of the 
dipole potential $U_{d}$ with distance $z$ for an internal angle of incidence of $\phi=42^{\circ}$ within a dielectric substrate of
dielectric constant $\epsilon_1=3$.  Also shown (dashed curve) are the variations of the new potential field-dependent 
$U_{g}^{f}$ and their sum $U_{total}$ (full curve).  The inset to the figure shows the atomic trajectory for an atom with initial velocity
components $v_{z}=-0.4$ ms$^{-1}$ and $v_{x}=0.6$ ms$^{-1}$. The initial position is $x_{0}=0; z_{0}=300$ nm.

\subsection*{\bf Figure 4}

Comparison of potentials and trajectories for the very low intensity case $(I=8.0\times 10^{2}$ Wm$^{-2})$ and the field-free
case. The main figure shows the variations of the dipole potential $U_{d}$ (dotted curve, almost coinciding with the horizontal axis), 
the field-dependent potential  
$U_{g}^{f}$ (full curve) and their sum $U_{total}$ (practically coinciding with the full curve).
The dashed curve shows the 
variations of the field-free potential $U_{g}^{0}$, exactly as in Fig. 1.
The inset compares the atomic trajectories arising from the two potentials for an atom with initial velocity components 
$v_{z}(0)=-0.067$ ms$^{-1}$ 
and $v_{x}(0)=0.4$ms$^{-1}$. 
The initial position is $x_{0}=0; z_{0}=1.13\mu$m. The trajectory shown by the full curve corresponds to the field free case.

\newpage

\pagebreak
\begin{figure}[tbh]
\includegraphics[totalheight=5in,angle=0]{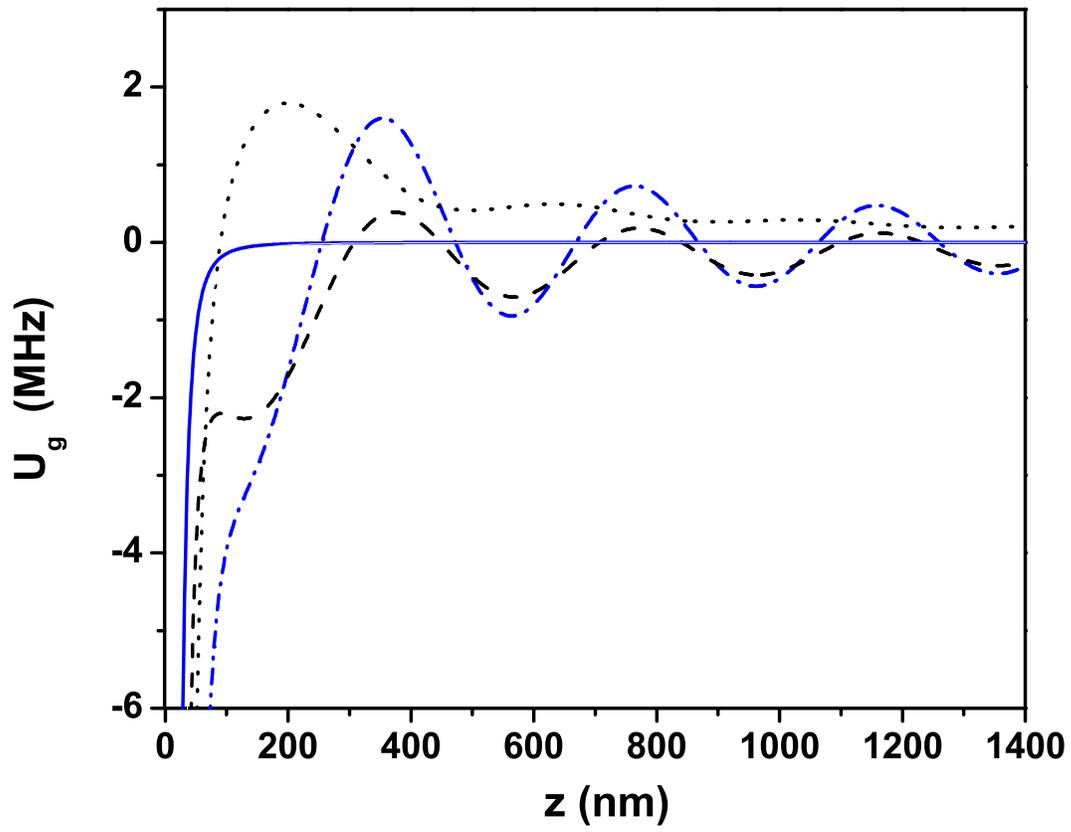}
\caption{Al-Amri  2006}
\label{fig1}
\end{figure}

\pagebreak
\begin{figure}[tbh]
\includegraphics[totalheight=5in,angle=0]{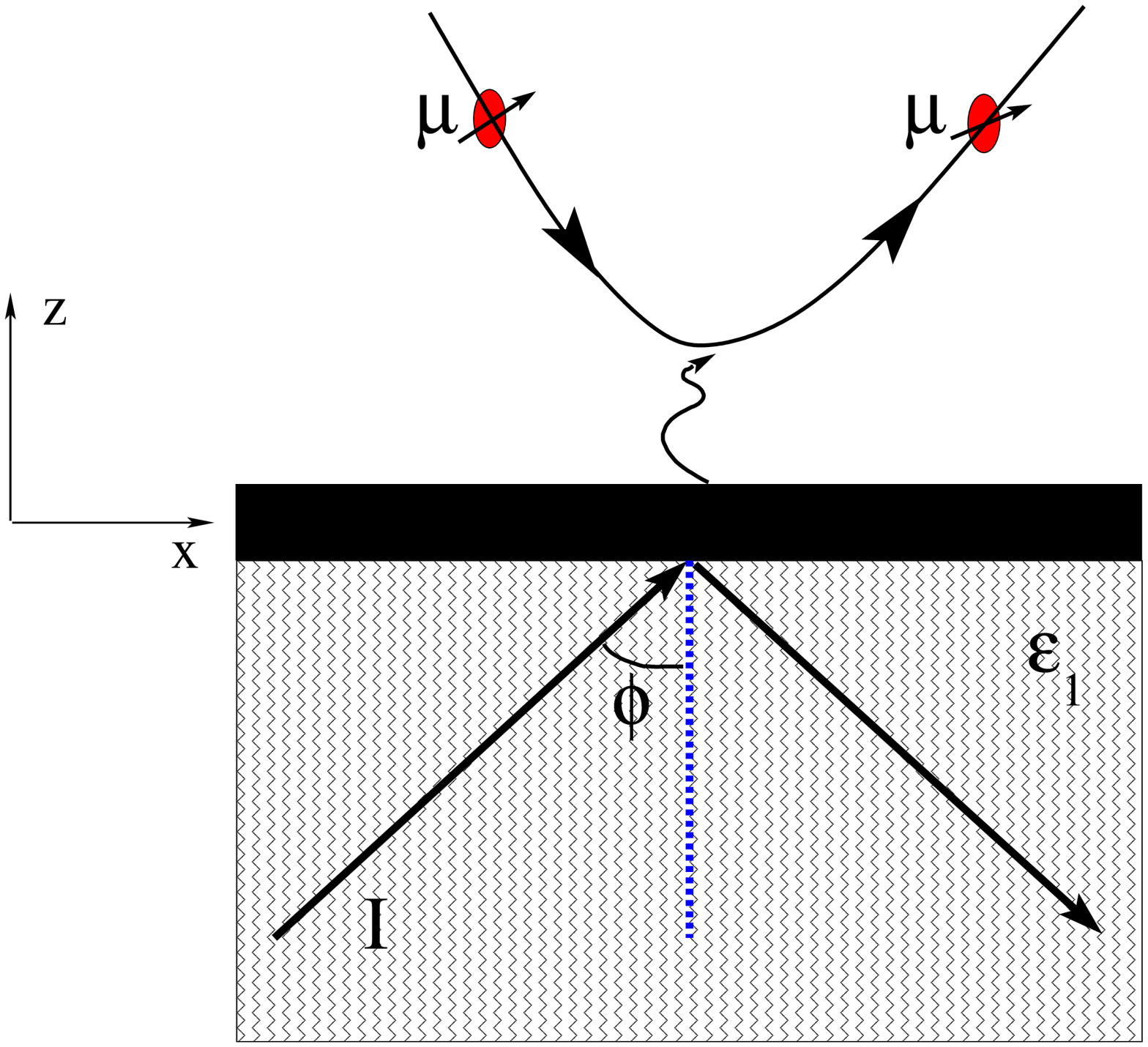}
\caption{Al-Amri  2006}
\label{fig2}
\end{figure}

\pagebreak
\begin{figure}[tbh]
\includegraphics[totalheight=5in,angle=0]{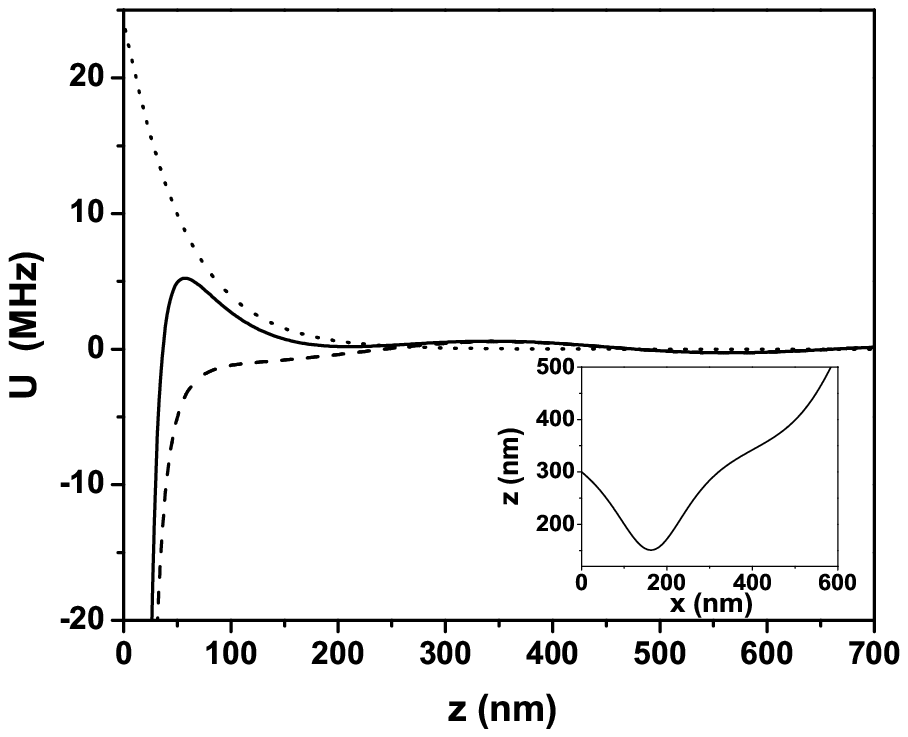}
\caption{Al-Amri  2006}
\label{fig3}
\end{figure}

\pagebreak
\begin{figure}[tbh]
\includegraphics[totalheight=5in,angle=0]{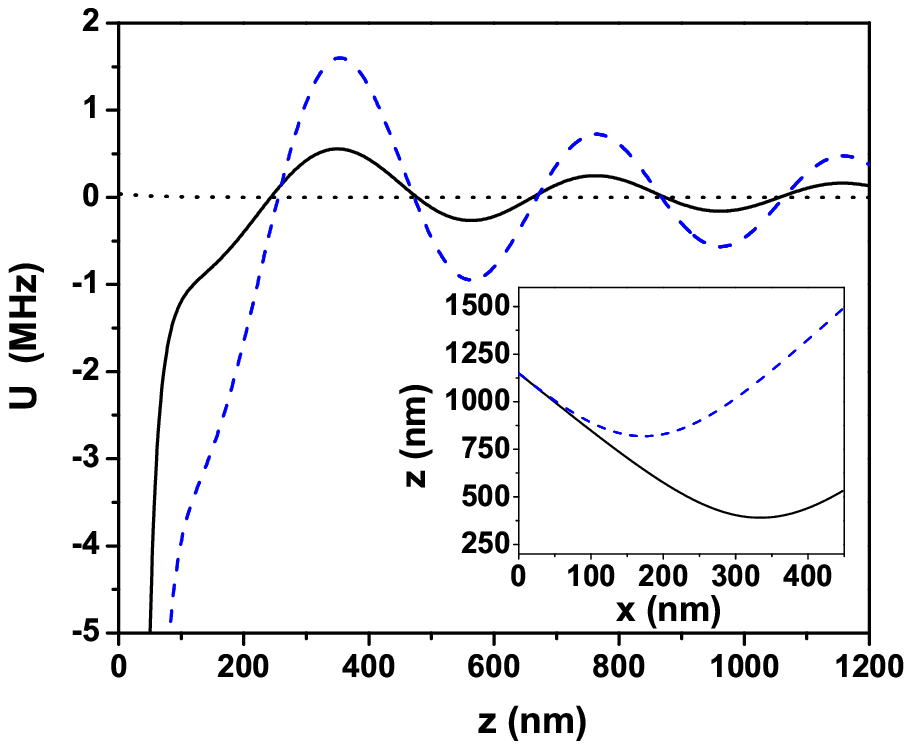}
\caption{Al-Amri  2006}
\label{fig4}
\end{figure}

\end{document}